\documentclass[12pt]{article}

\usepackage{amstext,amsmath,amssymb,amsfonts,bbm,slashed}
\usepackage[latin1]{inputenc}
\usepackage{epsfig}
\usepackage{hyperref}
\usepackage{amsthm}
\usepackage{subfigure}
\usepackage{color}
\usepackage{multirow}

\usepackage{hyperref}
\newtheorem{theorem}{Theorem}[section]

\newtheorem{definition}{Definition}[section]
\newtheorem{proposition}{Proposition}[section]

\newcommand{\beqa}{\begin{eqnarray}}
\newcommand{\eeqa}{\end{eqnarray}}

\newcommand{\bea}{\begin{eqnarray}}
\newcommand{\eea}{\end{eqnarray}}

\newcommand{\ZZ}{{\mathbb Z}}

\newcommand{\N}{{\mathbb N}}
\newcommand{\R}{{\mathbb R}}

\newcommand{\be}{\begin{equation}}
\newcommand{\ee}{\end{equation}}
\newcommand{\beq}{\begin{eqnarray}}
\newcommand{\eeq}{\end{eqnarray}}

\newcommand{\iid}{i.i.d.}
\newcommand{\mo}{m.o. }

\def\lsim{\
  \lower-2.0pt\vbox{\hbox{\rlap{$<$}\lower5.5pt\vbox{\hbox{$\sim$}}}}\ }
\def\gsim{\
  \lower-2.0pt\vbox{\hbox{\rlap{$>$}\lower5.5pt\vbox{\hbox{$\sim$}}}}\ }

\begin{document}

\begin{flushright}
\end{flushright}

\vspace{20pt}

\begin{center}

{\Large\bf 
The $1/N$ expansion of multi-orientable random tensor models}
\vspace{20pt}

St\'ephane Dartois${}^{a}$
\footnote{e-mail: stephane.dartois@ens-lyon.fr}, 
Vincent Rivasseau${}^{b,c}$
\footnote{e-mail: vincent.rivasseau@th.u-psud.fr} 
and
Adrian Tanasa${}^{a,d}$
\footnote{e-mail: adrian.tanasa@ens-lyon.org}



\vspace{10pt}

\begin{abstract}
\noindent
Multi-orientable  group field theory (GFT) has been introduced in 
\cite{mo},
as a quantum field theoretical simplification of GFT, 
which retains a larger class of tensor graphs 
than the colored one.
In this paper we define the associated multi-orientable identically
independent distributed multi-orientable tensor model
and we derive its $1/N$ expansion. In order to obtain this result, 
a partial classification of general tensor graphs is performed and 
the combinatorial notion of jacket is 
extended to the \mo graphs.
We prove that the leading sector is given, as in the case of colored models,
by the so-called melon graphs.
\end{abstract}

\end{center}




\setcounter{footnote}{0}

\section{Introduction.}
\label{Intro}
\renewcommand{\theequation}{\thesection.\arabic{equation}}
\setcounter{equation}{0}

The interface between combinatorics and theoretical physics is rapidly growing.
This interface has multiple aspects, such as 
the interplay of combinatorics (algebraic, analytic and so on)
with quantum field theory (QFT) \cite{combqft} (a typical example being 
the elegant Connes-Kreimer algebraic reformulation of the 
combinatorics of renormalization \cite{CK}), 
or the interplay of combinatorics (enumerative, bijective and so on) 
with statistical physics and integrable systems \cite{rev-df}.

This paper is situated at the interface between combinatorics 
and random tensor models, which naturally generalize random matrix models.
These models are interesting candidates for
a fundamental theory of quantum gravity \cite{tensortrack}, 
in relation to many other quantum gravity approaches, 
such as  dynamical triangulations \cite{dyntri}, loop quantum gravity \cite{bookrovelli},
non-commutative geometry \cite{bookconnes} or even
string theory \cite{bookstrings}.

Nevertheless, the combinatorics and the topology 
of these tensor models is rather involved. Progress came from a 
simplified colored version of these models \cite{cgft}, first proposed within 
the framework of group field theory (GFT) \cite{boulatov}
and then as an even more general theory of random tensors 
\cite{review,univ,uncolored}. Colored models discard a significant class of 
the initial graphs considered in earlier tensor models but retain triangulations of 
all piecewise linear manifolds in any dimension \cite{lost}. They admit a $1/N$ expansion when 
the size $N$ of the tensor becomes large \cite{1/N}. 
This expansion is a direct generalization of the matrix models one, 
for which the leading graphs are the celebrated \emph{planar} graphs 
(triangulating the ${\cal S}^2$ sphere). Its
leading sector is given by the so-called {\it melonic graphs}, which correspond to 
a \emph{particularly simple class} of triangulations of the ${\cal S}^D$ sphere, $D\ge 3$ being 
the dimension of space-time, or equivalently, the rank of the tensor.

This $1/N$ expansion for colored tensor models directly lead to the existence of a continuum phase transition 
and to the exact computation of its critical exponents \cite{bijection}. It also allowed to find associated renormalizable 
colored tensor field theories and tensor group field theories \cite{ren}.
\medskip

Recently, another QFT-inspired simplification of tensor models was defined in three dimensions - or, equivalently, for
rank-three tensors-, namely  multi-orientable (hereafter m.o.) models \cite{mo}. Defined initially again within the GFT framework,
they retain a larger class of tensor graphs than the colored models. In view of the many
applications of the $1/N$ expansion in the colored case, it is important to know whether
such \mo models also admit a 1/N expansion. In this paper we answer 
this question positively, proving that such a $1/N$ expansion exists
for independently identically distributed (hereafter \iid) \mo models.
We prove that the leading sector is given, as in the case of colored tensor models,
by the melonic graphs. In order to obtain these results, we give a partial classification of tensor graphs
and we generalize the combinatorial notion of jacket (embedded ribbon graphs defined in \cite{kra, 1/N}) 
to \mo graphs.

\medskip

The paper is structured as follows. In the next section 
we define the \iid tensor models in the \mo case.
The third section is devoted to our graph classification, with respect to edge-colorability 
and bipartitism. The following section gives some 
combinatorial and topological tools,
namely the generalization of the notion of tensor graph jackets 
and the definition of a generalized tensor graph degree.
The fifth section establishes the $1/N$ development. In the sixth section we 
prove 
that the leading sector of this expansion is again given by the melonic graphs,
using in particular Theorem \ref{nbp} which states that every non bipartite \mo graph has at least a non orientable jacket. The last section presents some concluding remarks and 
lists a few perspectives for future work within this \mo tensor framework.

\section{Definition of the model}
\label{sec:def}

We introduce in this section the \iid, \mo model and we study its symmetries. We consider a complex field $\phi$ taking values in a tensor product of three vector spaces $W=E_1\otimes E_2\otimes E_3$. 
Suppose we have a Hermitian or real scalar product in each $E_i$, $i=1,2,3$.
In analogy with complex matrices, we would like to introduce a Hermitian conjugation for the field $\phi$. 
Since in any orthonormal basis, the field has three indices, there is no canonical notion of transposition.
Nevertheless, giving a special role to the  indices $1$ and $3$, which we call \textsl{outer} indices, 
we can define a tensor field $\hat \phi$ \textit{via} the relation $\hat \phi_{kji}=\bar \phi_{ijk}$. 
In the limit of zero dimension for $E_2$ we recover the usual matrix model conjugation. 

A natural action for such a tensor model is:
\begin{align}
\label{action}
 S[\phi] &= S_0[\phi]+S_{int}[\phi], \notag \\
 S_0[\phi] &= \frac{1}{2}\sum _{i,j,k} {\hat\phi}_{kji} \phi_{ijk}, \  S_{int}[\phi] = \frac{\lambda}{4}\sum_{i,j,k,i',j',k'}\phi_{ijk}{\hat\phi}_{ij'k'}\phi_{k'ji'}{\hat\phi}_{i'j'k}.
\end{align}
Let us emphasize that the quadratic part of the action thus defined is positive. A second remark is that
the transposition of the outer indices in the definition of $\hat \phi$ implies that $\hat \phi \in E_3\otimes E_2 \otimes E_1$\footnote{We can remark
here that $\hat \phi $ could be interpreted as an object living in the dual of $W$, but 
the Hermitian product provides a canonical isomorphism between $W$ and its dual.}. The index contractions impose no additional constraints on the space $W$. 

Let us now suppose  that each vector space $E_i$ carries a representation of a Lie group $G_i$.
The tensor product $W$ then carries a natural representation 
of the group $G_1\times G_2 \times G_3$ namely 
the tensor product representation.
More explicitly, in a given basis the field transforms as
\begin{align}
 \phi'_{ijk}&=\rho_1(g_1)_{ii'}\rho_2(g_2)_{jj'}\rho(g_3)_{kk'}\phi_{i'j'k'},  \\
 \overline{\phi'_{ijk}}&=\overline{\rho_1(g_1)_{ii'}}\,\overline{\rho_2(g_2)_{jj'}}\,\overline{\rho(g_3)_{kk'}}\,\overline{\phi_{i'j'k'}},
\end{align}
where $\rho$ are the matrices of the group representations.
The corresponding transformation on  $\hat \phi$ is:
\begin{equation}
 \hat{\phi}'_{ijk} = \overline{\rho_3(g_3)_{ii'}}\,\overline{\rho_2(g_2)_{jj'}}\,\overline{\rho_1(g_1)_{kk'}}\,{\hat{\phi}}_{i'j'k'}.
\end{equation}

The natural invariance of the quadratic part 
in \eqref{action} 
is under the unitary 
groups $G_i=U(N_i)$.
However 
the interaction term restrains this invariance. 
The contraction of the second index of a $\phi$ (resp. $\hat \phi$) with the second 
index of a $\phi$ (resp. $\hat \phi$) field imposes $G_2$ to be an orthogonal group $O(N_2)$. The action of the \mo model is thus invariant
under  $U(N_1)\times O(N_2) \times U(N_3)$, and natural \mo models have a real rather than Hermitian scalar product on a real inner space $E_2$.
The spaces $E_1$ and $E_3$ could be either complex or real, in which case the invariance
is $O(N_1)\times O(N_2) \times O(N_3)$.

In the  limit of vanishing dimension for $E_2$ one recovers the matrix model action
\begin{align}
 S[M]=\frac{1}{2}Tr[M^{\mathcal{y}}M]+\frac{\lambda}{4}Tr[(M^{\mathcal{y}}M)^2].
\end{align}
This provides a natural interpolation 
between random tensors and random matrices models.

\begin{figure}
\begin{center}
\includegraphics[scale=0.38]{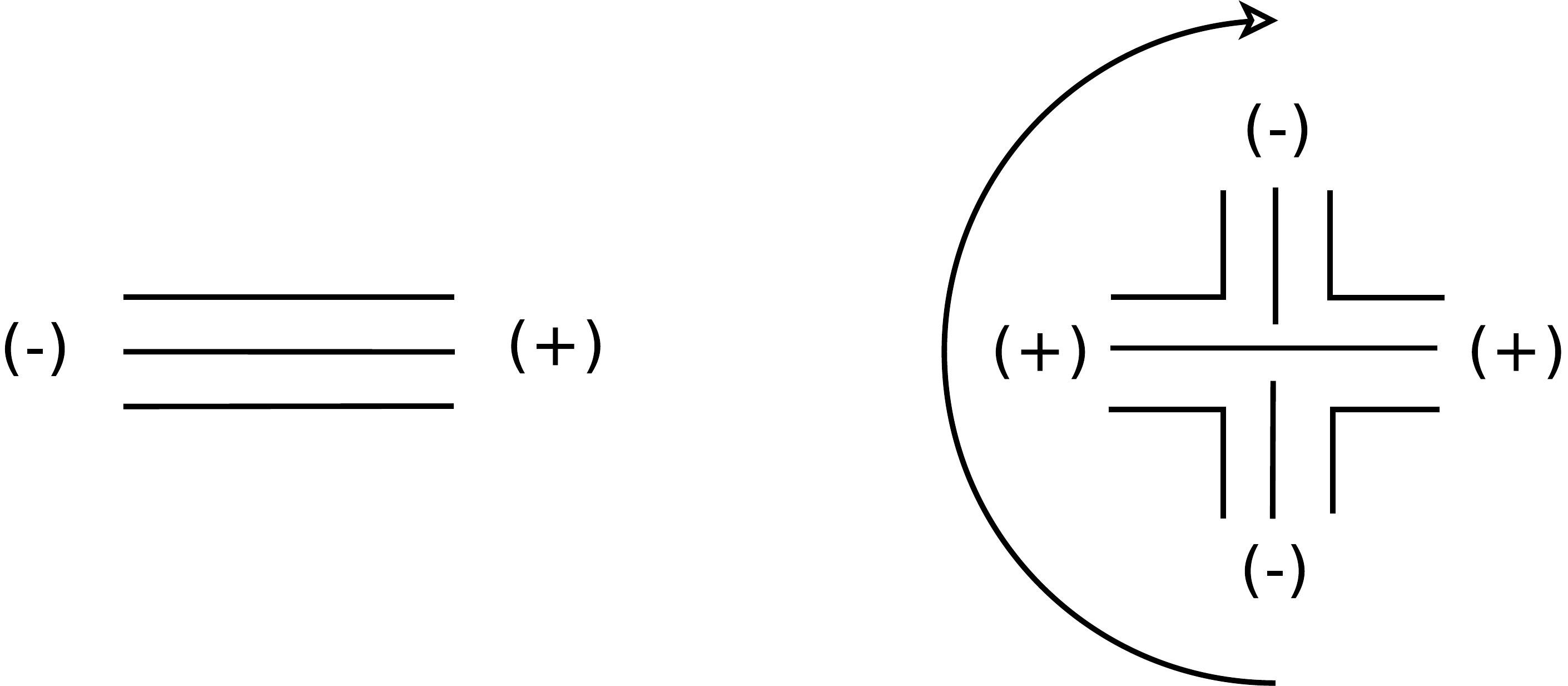}
\caption{Propagator and vertex of the \mo model}\label{propagator} 
\end{center}
\end{figure}

The Feynman graphs associated to the action \eqref{action} are built from 
the propagator and the vertex of Fig. \ref{propagator}. 

These Feynman graphs are called stranded graphs. The strands represent the indices of the tensor field (analogous to
ribbon boundaries in the case of the ribbon graphs of matrix models). The $(+)$ and $(-)$ signs appearing at the vertex represent (respectively) the $\phi$ and $\hat \phi$ fields occurring in the
interaction term of the action. The propagator has no twist on the strands (as a consequence of the form of the quadratic part of the action). The order of the strands (representing indices) at the vertex is induced by the choice of a cyclic orientation around the vertex.
As in  \cite{mo}, we call these Feynman graphs \mo graphs. They are stranded graphs for which there exists a labeling
with $(+)$ and $(-)$ at the vertices such that an edge always connects a $(+)$ sign with a $(-)$ sign, hence edges are \emph{oriented}. 

\medskip

Let us also recall here the definition of the colored \iid model.
One has a quadruplet of complex fields 
$\left(\phi^0,\phi^1,\phi^2,\phi^3\right)$; the action is given by:
\begin{align}
\label{color}
 S[\{\phi^i\}] &= S_f[\{\phi^i\}]+S_{int}[\{\phi^i\}] \notag \\
 S_f[\{\phi^i\}]&=\frac12 \sum_{p=0}^{3}\sum_{ijk} \overline{\phi^p_{ijk}}\phi^p_{ijk}\notag \\
 S_{int}[\{\phi^i\}]&=\frac{\lambda}{4}\sum_{i,j,k,i',j',k'}\phi^0_{ijk}\phi^1_{i'j'k}\phi^2_{i'jk'}\phi^3_{k'j'k}
+\mbox{ c. c.},
\end{align}
where by $\mbox{ c. c.}$ we mean the complex conjugated term.
We refer to the indices $0,\ldots, 3$ as to color indices. 
Moreover, let us mention that this action implies that the faces of the Feynman graphs of this model
have always exactly two (alternating) colors.

\medskip

Finally we equate the dimensions of the three spaces $N_1=N_2=N_3=N$ in order to perform a $1/N$ expansion\footnote{In GFT a stronger restriction $E_1=E_2=E_3$ is required, 
because models such as the Boulatov model \cite{boulatov}  
 include a projector which 
averages over a common action on these spaces.}.
The GFT case corresponds to  these three spaces being 
finite dimensional truncations of $L_2(G)$
where $G$ is a compact Lie group. 
Indeed, a compact Lie group has a unique Haar measure and a unique 
corresponding infinite dimensional space of square integrable function $L_2(G)$. 
A finite dimensional truncation $E_i=L_2^\Lambda(G)$ of 
this space can be obtained by retaining Fourier modes (or characters in the Peter-Weyl expansion) up 
to some cutoff
  $\Lambda$. This space is finite dimensional and is Hilbertian (since it inherits 
the Hermitian product of $L_2(G)$). 
For instance, in the simplest case of the group $U(1)$, the truncation which retains
modes $n \in \ZZ$ with $\vert n\vert \le \Lambda$ correspond to a space 
$L_2^{\Lambda}(U(1))$ of dimension $N = 2\Lambda +1$.

 \section{Classification of Feynman graphs}
\label{sec:class}

The partition function of the \mo model is given by:
\begin{align}
\label{partitionfunction}
Z(\lambda)& = \int \mathcal{D}\phi e^{-S[\phi]} = e^{-F(\lambda)}.
\end{align}
As usual the quantity of interest is its logarithm (up to some normalization),
\begin{align}
\label{freeenergy}
F(\lambda)& = -\ln [Z(\lambda)] = \sum_{\mathcal{G}} \frac{1}{s(\mathcal{G})} A^{}_{}(\mathcal{G}),
\end{align}
where the sum runs over connected vacuum \mo Feynman graphs $\mathcal{G}$. From now on 
we consider only vacuum connected graphs.

Let us emphasize that the Feynman graphs of tensor models
are from a mathematical point of view CW-complexes. Nevertheless, since we do not allow twists on the propagators of these models,
there is a one-to-one correspondence between these CW-complexes and graphs.
We can thus apply to these objects several results known from graph theory\footnote{Let us point out what we mean by graph here. In general we are interested
in  Feynman graphs used to label amplitudes of a (group) field theory. Forgetting the strands, these Feynman graphs are, from a mathematical point of view, multi-graphs (i.e. allowing multiple edges between vertices) and pseudo-graphs (tadpole graphs, in which there are edges with both end on the same vertex).} .
We call \emph{four-edge colorable} a graph for which the edge chromatic number is equal to four.

\begin{figure}
\begin{center}
\includegraphics[scale=0.99]{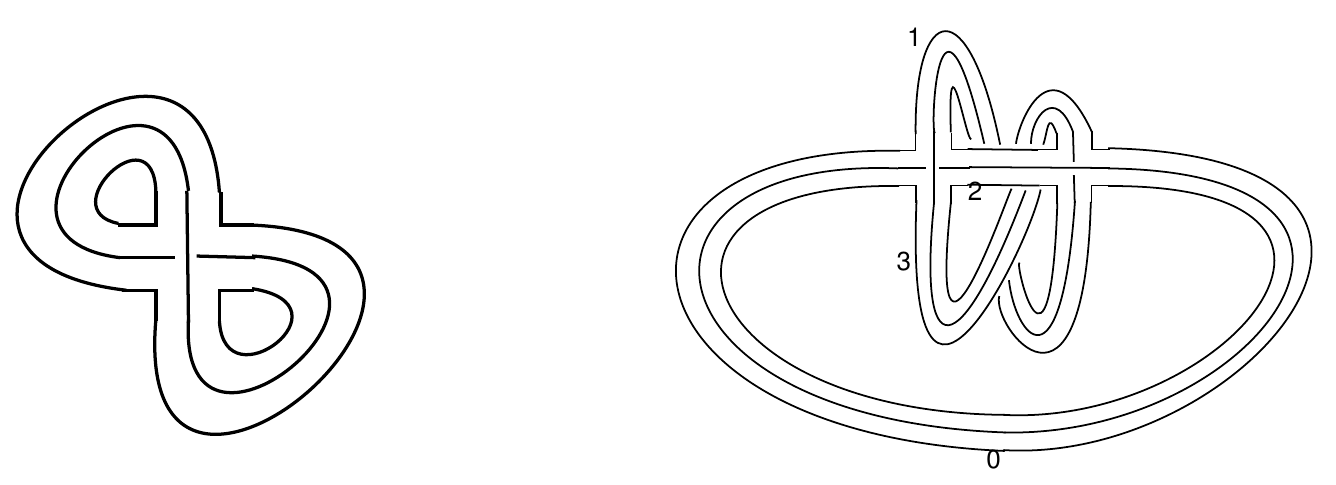}
\caption{\label{planartadtwistsun} On the left one can see the planar double tadpole as an example of a \mo graph which is not colorable. On the right is pictured the "twisted sunshine" as an example of a \mo graph which is 4-edge
colorable but does not occur in colorable models.}
\end{center}
\end{figure}

\begin{proposition}
The set of Feynman graphs generated by the colored action \eqref{color} 
is a strict subset of the set of Feynman graphs generated by the \mo action \eqref{action}.
\end{proposition}
\proof  As already mentioned above, the action \eqref{color} generates Feynman graphs
with exactly two colors on each face. 
The \mo  action \eqref{action} generates graphs which are four-edge colorable, 
but with faces having more than two colors in any coloring choice. 
 An example of such a graph is given in Fig. \ref{planartadtwistsun} right. 
Hence, the \mo graphs form a strictly larger class than the colored graphs. \qed

\begin{figure}[h]
  \begin{center}
   \includegraphics[scale=0.55]{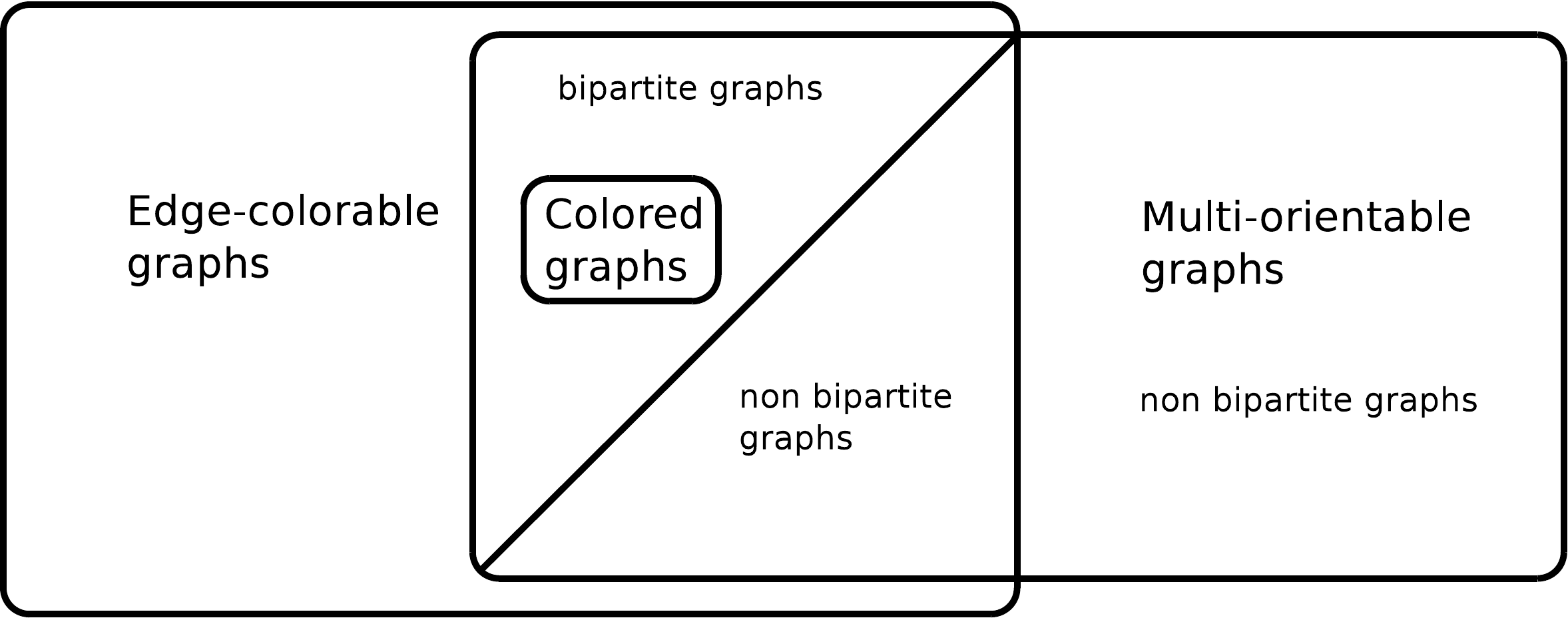}
   \caption{\label{classification} Tensor graphs classification.}
  \end{center}
\end{figure}

\medskip
Since we deal in this paper with graphs for which the vertices have maximal 
incidence number equal to four, one has:

\begin{proposition} 
A bipartite graph is four-edge colorable.
\end{proposition}

\proof The proof is a direct consequence of  Theorem $2$ of Chp. $12$ of  \cite{book-berge},  which states that the 
chromatic number of a bipartite graph\footnote{The theorem is stated for bipartite multigraphs.}
is equal to the maximum degree of its vertices. In our case all vertices have constant degree equal to four. Then all bipartite graphs appearing in the multi-orientable model are four-edge colorable.  \qed

\begin{figure}[htb]
\begin{center}
\includegraphics[scale=0.15]{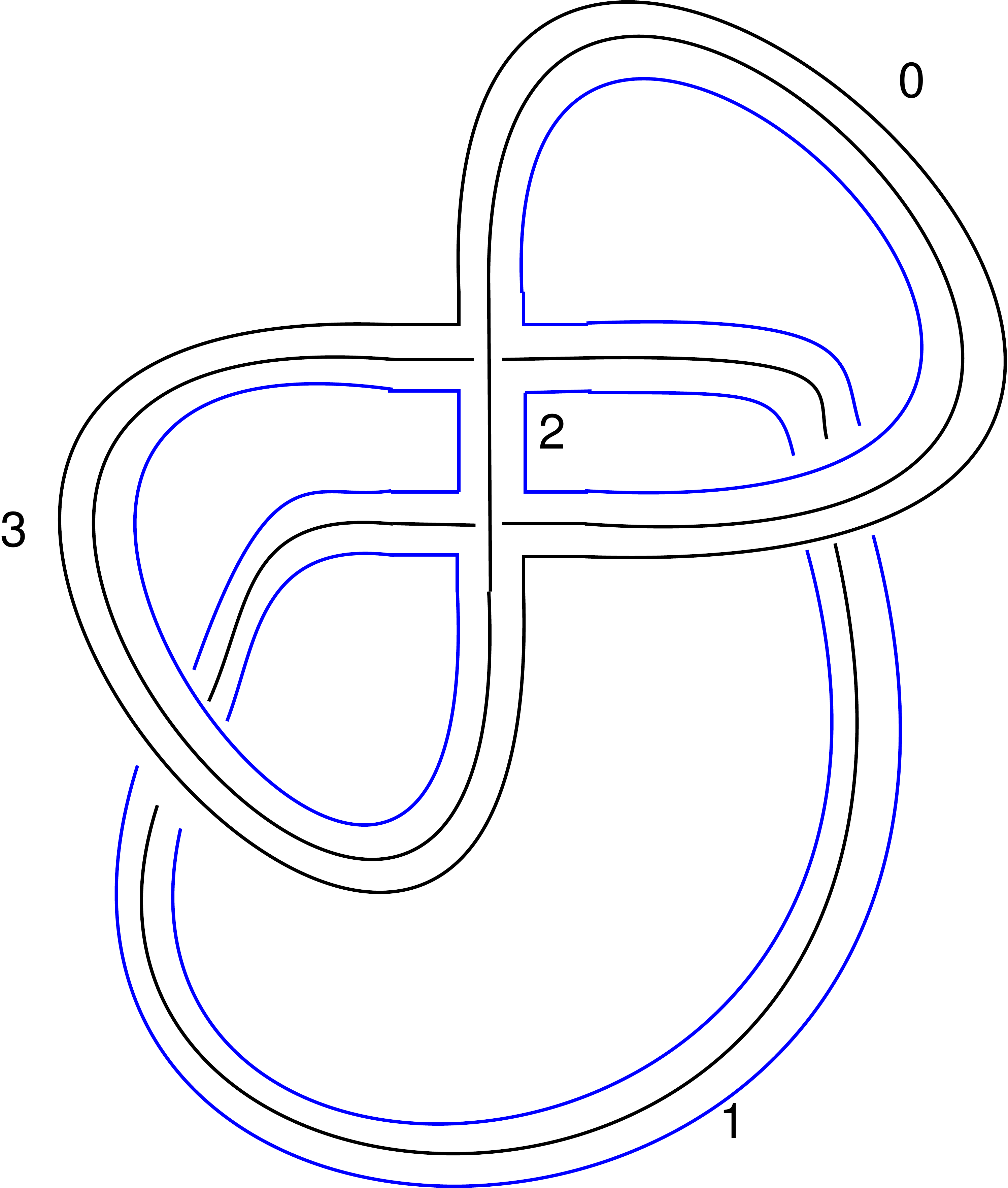}
\caption{\label{ectadface} Example of a graph with a tadface
which is edge-colorable. The tadface line is shown in blue. 
The numbers $0$, $1$, $2$, $3$ are the color labels of the edges.}
 \end{center}
\end{figure}

\begin{figure}[htb]
 \begin{center}
 \includegraphics[scale=0.10]{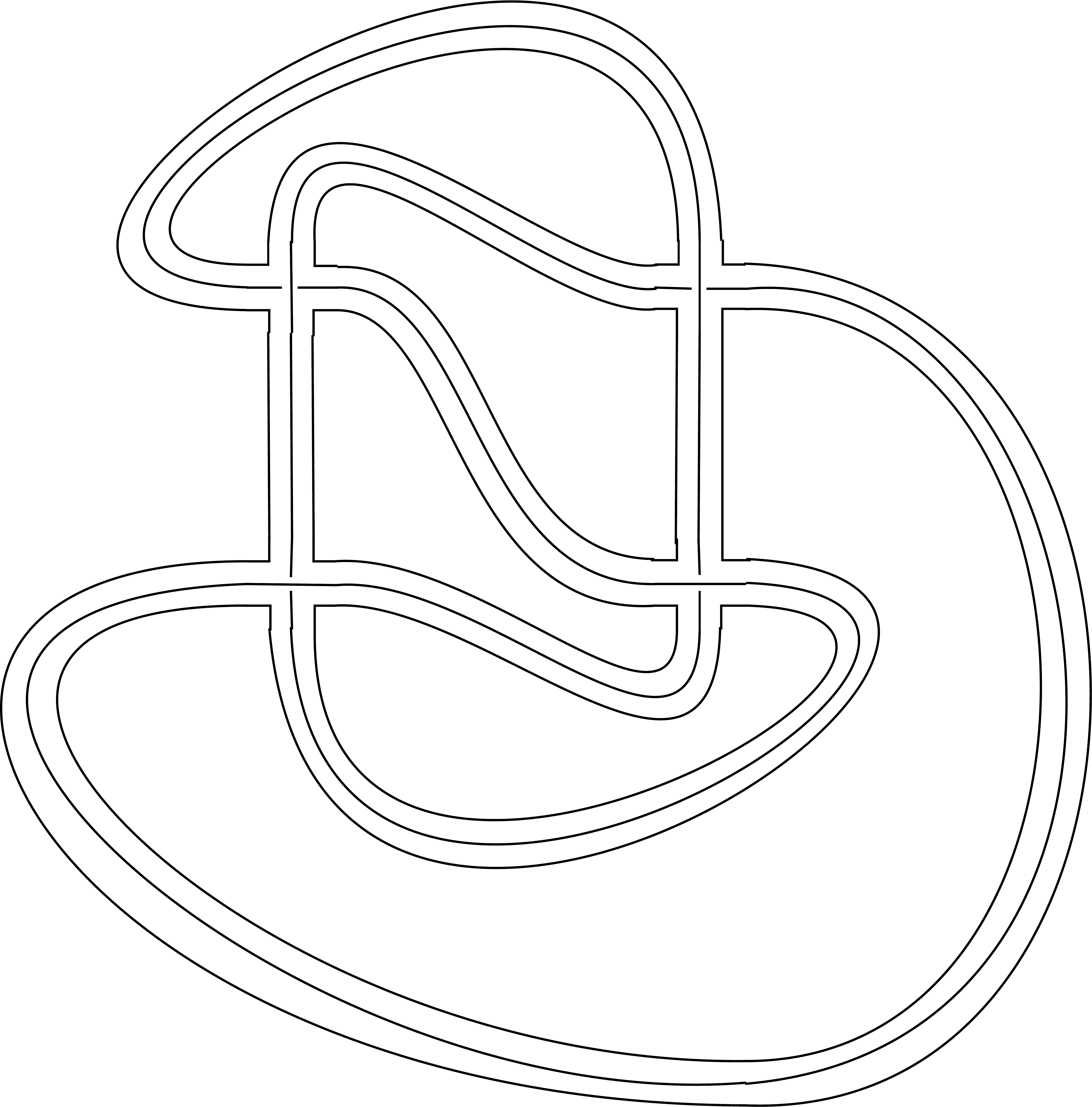}
 \caption{\label{moecnonbip} A 4-edge colorable \mo graph which is not bipartite.}
 \end{center}
\end{figure}

\begin{figure}[htb]
\begin{center}
\rotatebox{90}{\includegraphics[scale=0.5]{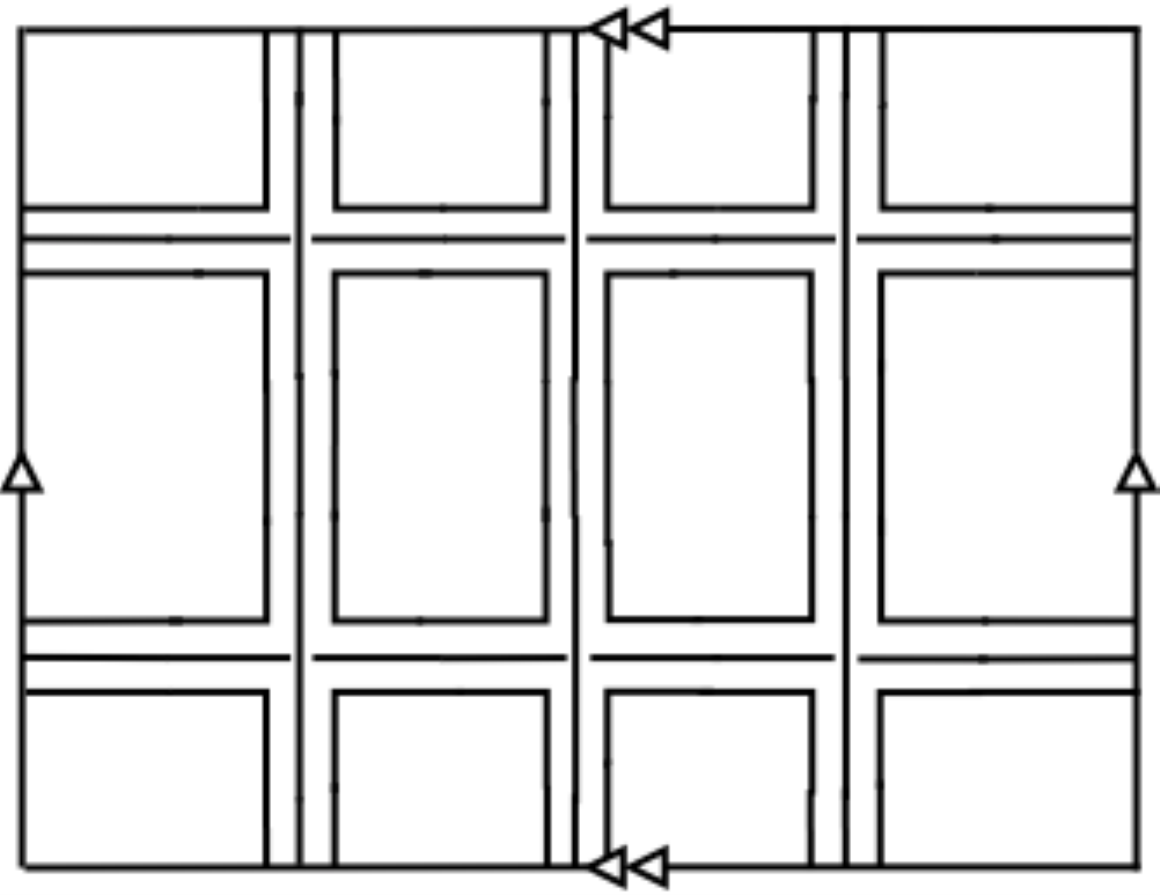}}
\caption{\label{nomonotadface} A graph without tadface which is not m.o. 
Edges of the box are identified so that the graph is drawn on the torus.}
\end{center}
\end{figure}

We then get the classification of Fig.  \ref{classification}. 
Let us also give some examples of different graphs appearing in each subset of this classification,
see Fig. \ref{ectadface}, \ref{planartadtwistsun}, \ref{moecnonbip}, \ref{nomonotadface}.
An important example is the one of Fig. \ref{nomonotadface} - a graph without tadface
(a tadface being a face ``going'' several times through the same edge) 
 which is not \mo. Let us recall that the condition of multi-orientability discards tadfaces (see Theorem $3.1$ of \cite{mo}).
Fig. \ref{planartadtwistsun} left gives an example of a \mo graph which is non bipartite (and non colorable). Fig.  \ref{planartadtwistsun} right gives an example of
a graph which is 4-edge colorable and multi-orientable but not colorable in the sense of action \eqref{color}. In fact one can check there are two faces with four colors, hence the graph cannot be generated by the action of the colored tensor model.

\medskip

One last example is the graph of Fig.\ref{nomonotadface}. It can be  drawn on the torus 
(the side of the box having the same types of arrows being identified). This graph has no tadface and yet is not multi-orientable. As checked later 
it has no well-defined jackets.



\section{Combinatorial and topological tools}
\renewcommand{\theequation}{\thesection.\arabic{equation}}
\setcounter{equation}{0}
\label{sec:tools}

In the colored case the $1/N$ expansion  \cite{1/N} relies on the notion of jackets.
Jackets are ribbon graphs\footnote{Recall that ribbon graphs, are made of vertices homeomorphic to disc and edges (maybe twisted) connecting them.
Perturbation theory for random Wishart (i.e. possibly rectangular) matrix models is indexed by such ribbon graphs.} 
associated to the cycle of colors up to orientation.
The colored $1/N$ expansion is indexed by an integer called the \emph{degree}
which is the sum of the genera of all jackets.
In particular in dimension 3 colored tensor graphs have three jackets which define
three different Hegaard splitting of the dual of the underlying 
(pseudo)manifold \cite{ryan}.

In the multi-orientable case we want to implement a similar $1/N$ expansion, hence we need to generalize the notion
of jackets. We remark that six strands meet  at any vertex $v$. In the most general case there is no way to split them into
three pairs $a_v, b_v, c_v$ in a coherent way throughout the graph, namely in such a way that
any face is made out only of strands of the \emph{same type}, $a$, $b$ or $c$ (see Fig.  \ref{nomonotadface} 
for an example of a stranded graph where such a splitting is impossible).

\begin{figure}[htb]
\begin{center}
 \includegraphics[scale=0.32]{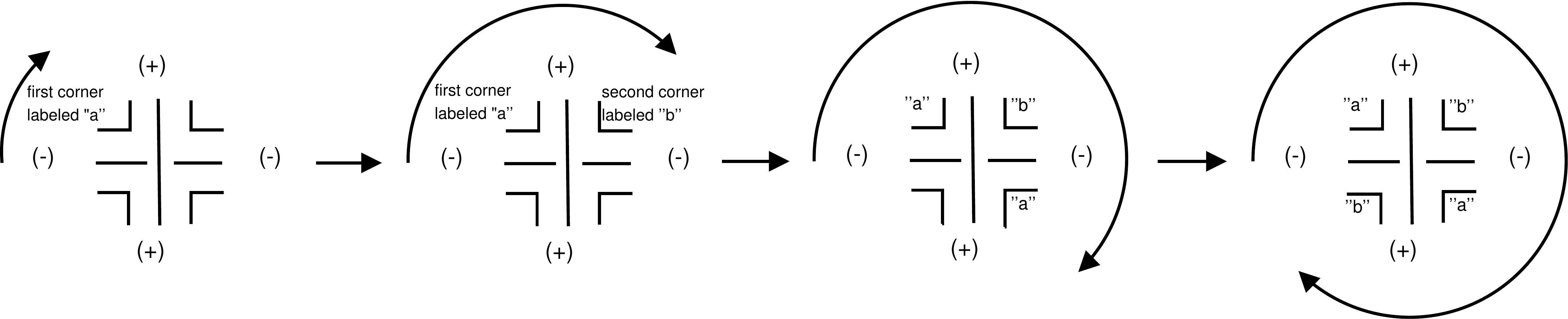}
 \caption{\label{labeling}Labeling procedure of a \mo vertex.}
\end{center}
\end{figure}

\begin{figure}[htb]
\begin{center}
\includegraphics[scale=0.6]{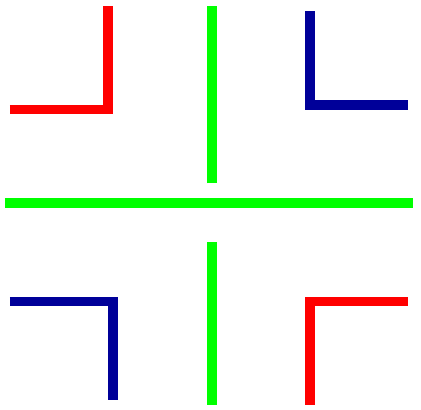}
\caption{\label{oppositestrands} This figure shows the three pairs of opposite corner strands at a vertex.} 
\end{center} 
\end{figure}

In the \mo case, such a coherent splitting is possible. Indeed at each vertex the inner pair of strands (those acting 
on the $E_2$ space) is unique and well-defined. Let us say that this pair has type $c$. This pair is coherent, that is
throughout the graph any face containing an inner strand is made of inner strands only. But we can also split at each 
vertex the four outer strands, or ``corner strands", into two coherent pairs, of  \emph{opposite strands}.
Consider indeed a vertex and turn clockwise around it starting at the inner strand of a $\hat \phi$ field, as shown on Fig. \ref{labeling}.
We call the first and third corner strands we meet corners of type $a$, and the second and fourth corner strands of type $b$. 
Thus the opposite strands at any vertex all have the same label (see Fig.  \ref{oppositestrands}). Since in a multi-orientable graph all vertices 
have canonical orientation, any $a$ (respectively $b$ type) type corner of a vertex always 
connects to an $a$ (resp. $b$) type corner of another vertex, hence faces made out of outer strands are made either 
entirely of $a$ strands or entirely of $b$ strands. This can be understood also because the $a$ strand correspond to the space $E_3$
in $W$ and the $b$ strands to the space $E_1$. The theory is consistent even for $E_1 \ne E_3$, hence cannot branch together
strands of type $a$ with strands of type $b$. Remark also that each edge contains three strands of the three different types, $a$, $b$ and $c$.




Strand type allows to define the three \emph{jackets} $\bar a$, $\bar b $ and $\bar c$
of an \mo tensor graph:
\begin{definition}
A jacket of an \mo graph is the graph made by excluding one type of strands throughout the graph. 
The \emph{outer} jacket $\bar c$ is made of all outer strands, or equivalently excludes the inner strands; 
jacket $\bar a$ excludes all strands of type $a$ and jacket $\bar b$ excludes all strands of type $b$.
\end{definition}

\begin{figure}[htb]
\begin{center}
\includegraphics[scale=0.15]{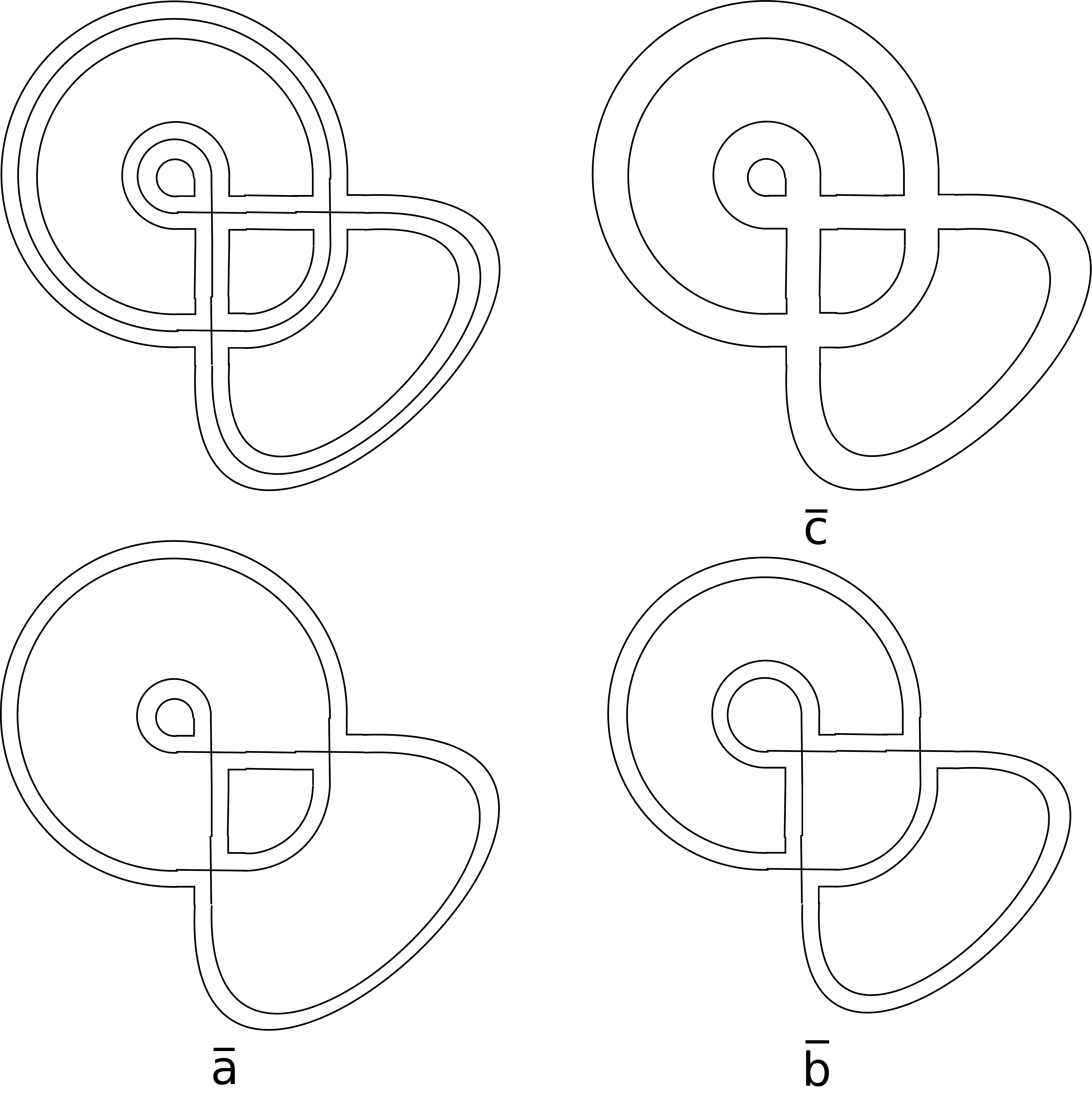}
\caption{\label{jacketexample} A \mo graph with its three jackets $\bar a$, $\bar b$, $\bar c$.}
\end{center}

\end{figure}

Fig. \ref{jacketexample} gives an example of a \mo graph with its three jackets.
The rest of this section is devoted to the following Proposition \ref{jac}:

\begin{proposition}
\label{jac}
Any jacket of a \mo graph is a (connected vacuum) ribbon graph (with uniform degree 4 at each vertex). 
\end{proposition}

\begin{figure}[htb]
\begin{center}
 \includegraphics[scale=0.18]{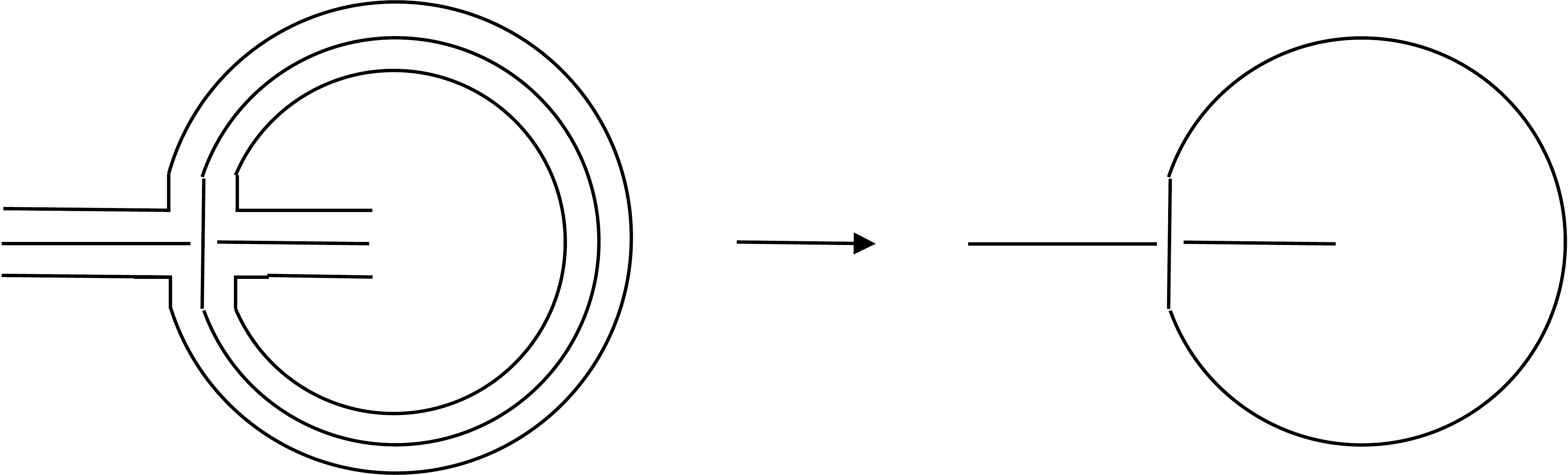}
 \caption{\label{algotadface} Deleting a pair of opposite corner strands in this tadpole (which has tadfaces),
does not lead to a 2-stranded graph.}
\end{center}
\end{figure}

Let us emphasize that Proposition \ref{jac} does not hold for general non-\mo graphs, 
for example in the case of a graph with a tadface (see Fig.\ref{algotadface}).

\proof The three jackets 
represent three random matrix graphs, restricted to three particular Wieshart random matrix models, respectively on spaces $E_1 \otimes E_3$, $E_1 \otimes E_2$
and $E_2 \otimes E_3$. Hence jackets are two-stranded graphs and in fact ribbon graphs. They are connected because the initial graph
itself was connected and because each edge contains three strands of three different types.  
\qed

\begin{figure}[ht]
\begin{center}
\includegraphics[scale=0.5]{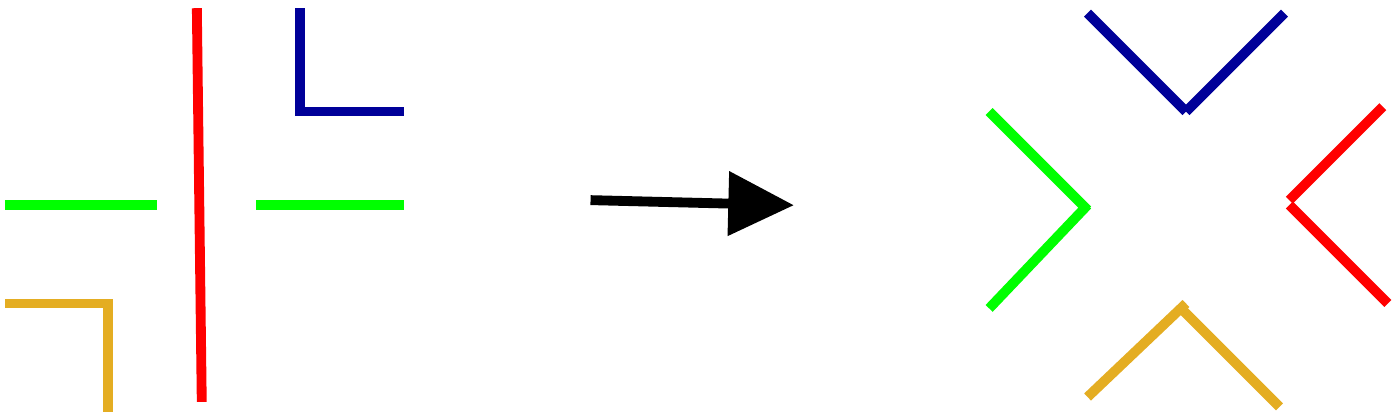}
\caption{\label{untwist} Untwisting a vertex. The strands are colored in order to clearly make the correspondence between the strands
before and after the untwisting of the vertex.}
 \end{center}
\end{figure}

Let us now give more explanations on this issue. \textsl{A priori} a 2-stranded graph may not be a ribbon graph because vertices may be \emph{twisted}.
Remark that this cannot happen for the outer jacket but may happen for the two others, see Fig. \ref{jacketexample}.
In that case we just have to untwist the vertices coherently  throughout  the whole graph (\textsl{i.e.} keeping the same set of faces and the same adjacency relations),
as shown in Fig. \ref{untwist}. This untwisting procedure can be performed by labelling the strands, then cutting the edges around the twisted vertex,
untwisting the vertex and then reconnecting the strands respecting the labeling of the strands. This last step may twist the 
new edges, but since these twists are introduced locally around the vertex the procedure can be continued coherently on all the vertices of the graph,
resulting in as much twists as necessary along the edges. 

Recall that ribbon graphs can represent either orientable or non-orientable surfaces.
Since the adjacency relation are invariant under the untwisting procedure, we can compute their Euler characteristic directly on the jacket 
independently of whether their vertices are twisted or not. Recall again that the Euler characteristic is related to the non-orientable genus $k$
through $\chi(\mathcal{J})=v-e+f=2-k$, where $k$ is the non-orientable genus, $v$ is the number of vertices, $e$ the number of edges and $f$ the number of faces. When the surface is orientable, $k$ is even and equal to twice the usual orientable
genus $g$ (so that we recover the usual relation $\chi(\mathcal{J})=2-2g$).

The \emph{degree} of a \mo graph 
$\mathcal{G}$ is given by:

\begin{definition}
Given a multi-orientable graph $\mathcal{G}$, its degree $\varpi(\mathcal{G})$ is defined by 
$$\varpi(\mathcal{G})=\sum_{\mathcal{J}} \frac{k_{\mathcal{J}}}{2},$$ 
the sum over $J$ running over the three jackets  of $\mathcal{G}$.
\end{definition}
In the colored case, all jackets are orientable and this formula gives back the colored degree. 
We also notice that the degree is a positive integer or half-integer.

\section{1/N expansion of \mo i.i.d. model}
\label{sec:exp}

Let us now organize the series \eqref{freeenergy}   according to powers of $N$,
$N$ being the size of the tensor. Since each face corresponds to a closed cycle of Kronecker $\delta$ functions, each face contributes with a factor $N$, where $N$ is the dimension of $E_1, E_2$ and $E_3$.
\begin{align}
 \label{1stamplitude}
 A^{}_{}(\mathcal{G})= \lambda^{v_{\mathcal{G}}} (k_N)^{-v_{\mathcal{G}}} N^{f_{\mathcal{G}}},
\end{align}
where $v_{\mathcal{G}}$ is the number of vertices of $\mathcal{G}$, $f_{\mathcal{G}}$ is the number of faces of $\mathcal{G}$ and $k_N$ is a rescaling constant.
 We choose this rescaling $k_N$ to get the same divergence degree
for the leading graphs at any order.
We first count the faces of a general graph $\mathcal{G}$ using the  jackets $\mathcal{J}$ of $\mathcal{G}$. 
From Euler characteristic formula, one has:
\begin{align}
 \label{facejacket}
f_{\mathcal{J}}= e_{\mathcal{J}}-v_{\mathcal{J}}-k_{\mathcal{J}}+2.
\end{align}
Since each jacket of $\mathcal{G}$  is a  connected vacuum ribbon graph, one has: $e_{\mathcal{J}} = 2v_{\mathcal{J}}$.
Let us recall here that the numbers of vertices (resp. edges) of a jacket 
$\mathcal{J}$ of $\mathcal{G}$ are the same than the numbers of vertices (resp. edges) of $\mathcal{G}$. 
Since each graph has three jackets and each face of a graph occurs in two jackets,  
summing \eqref{facejacket} over all the jackets of $\mathcal{G}$ leads to:
\begin{align}
 \label{facegraph}
f_{\mathcal{G}}&=\frac{3}{2}v_{\mathcal{G}}+3-\sum_{\mathcal{J} \subset \mathcal{G}} \frac{k_{\mathcal{J}}}{2} 
=\frac{3}{2}v_{\mathcal{G}}+3-\varpi(\mathcal{G}).
\end{align}
The amplitude rewrites as:
\begin{align}
 \label{2ndamplitude}
 A^{}_{}(\mathcal{G})= \lambda^{v_{\mathcal{G}}} (k_N)^{-v_{\mathcal{G}}} 
N^{\frac{3}{2}v_{\mathcal{G}}+3-\varpi(\mathcal{G})}.
\end{align}
To to get the same divergence degree
for the leading graphs at any order, we choose the scaling constant 
$k_N$
as being equal to $N^{\frac{3}{2}}$. The amplitude finally writes as:
\begin{align}
 \label{3rdamplitude}
A^{}(\mathcal{G})= \lambda^{v_{\mathcal{G}}} 
N^{3-\varpi(\mathcal{G})}.
\end{align}
Thus using the expression \eqref{3rdamplitude} for the amplitude we can rewrite the free energy as a formal series in $1/N$:
\begin{align}
 \label{newserie}
F(\lambda,N)&= 
\sum_{\varpi\in \N/2}C^{[\varpi]}(\lambda)N^{3-\varpi} \\
C^{[\varpi]}(\lambda)&=\sum_{\mathcal{G}, \varpi(\mathcal{G})=\varpi} \frac{1}{s(\mathcal{G})}\lambda^{v_{\mathcal{G}}}.
\end{align}

\section{Leading graphs }
\label{sec:melon}
\renewcommand{\theequation}{\thesection.\arabic{equation}}
\setcounter{equation}{0}

\begin{figure}[htb]
\begin{center}
\includegraphics[scale=0.25]{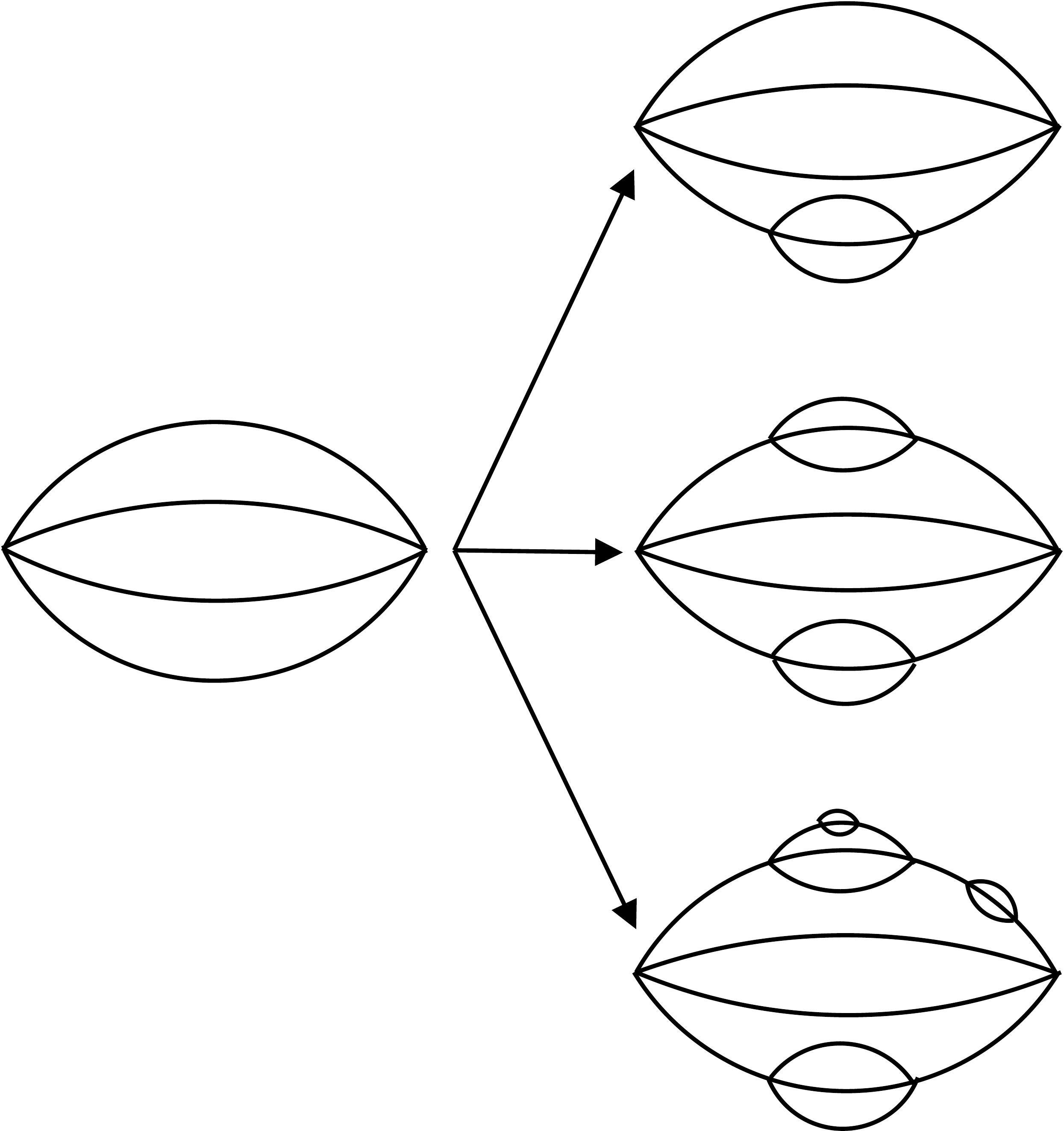}
\caption{\label{insertion} Some of the members of the melonic family.}
\end{center}
\end{figure}

As seen on equation \eqref{newserie}, the graphs which lead the $1/N$ expansion are those satisfying the relation $\varpi=0$.
Let us now  identify them from a combinatorial point of view. In the colored case 
the leading graphs, those of degree 0, called melonic graphs \cite{bijection}, are obtained by recursive insertions 
of the fundamental melonic two-point function on any line, starting from the fundamental elementary vaccum melon  (see Fig. \ref{insertion}). 
This fundamental vaccum melon has two vertices and four internal lines 
and the fundamental melonic two-point function has two vertices and three internal lines and two external legs of a fixed 
color.
Melon graphs can be mapped to $3-$ary trees, 
and counted exactly (by Catalan numbers).

As a first step let us compute the degree of three different \mo graphs, the ``double tadpole", 
the ``twisted sunshine", and the elementary melon.

Consider the jackets of the double tadpole of Fig.  \ref{planartadtwistsun} left.  
Its outer jacket is planar hence has genus $k_1=2g_1=0$. 
The second jacket is the one obtained by the elimination of the two faces of length one. This gives a ribbon graph representation of 
real projective plane $\R P^2$ with non-orientable genus $k_2=1$. It can be seen in two ways that the associated surface is non orientable. 
The first one is to directly glue a disc $D^2$ on the external face of the two stranded graph, to get a M\"obius band. Then gluing
a disc on the remaining face we get $\R P^2$. Otherwise we can untwist the vertex, obtaining a ribbon graph with one twist on each edge. This means that
we can find a path on the ribbon graph such that the local orientation is reversed after one turn. Finally the third jacket is equivalent to the
first one and represents the sphere. Thus the generalized degree of the double tadpole is 
$\varpi=\frac{1}{2}$.

The twisted sunshine is a bipartite 4-edge colorable graph. Its outer jacket is orientable 
(as is always the case for the outer jacket), and it has genus $g_1=1$. The two remaining jackets are isomorphic and represent 
the real projective plane. They have a non-orientable genus $k_2=k_3=1$. Thus the degree of the twisted sunshine 
is $\varpi=2$.

The elementary melon \cite{bijection} is even simpler. Its first jacket, the outer one, is 
planar. The two others also have genus zero (as follows directly from their Euler characteristic). Hence the 
generalized degree of this graph is $\varpi=0$. This is the first example of a graph leading the $1/N$ expansion of the \mo model. 

Let us now  determine the class of graphs which are leading in the $1/N$ expansion. 

\begin{theorem}
\label{nbp}
Non bipartite \mo graphs contain at least one non-orientable jacket and thus are of degree $\varpi \ge \frac{1}{2}$.
\end{theorem}

\proof Let $\mathcal{G}$ be a non bipartite multi-orientable graph. Then there is at least one odd cycle in $\mathcal{G}$ 
(see, for example, \cite{book-berge})
with $n$ vertices, $n$ being an odd integer.
\begin{figure}[ht]
\begin{minipage}[b]{0.45\linewidth}
\centering
\includegraphics[scale=0.34]{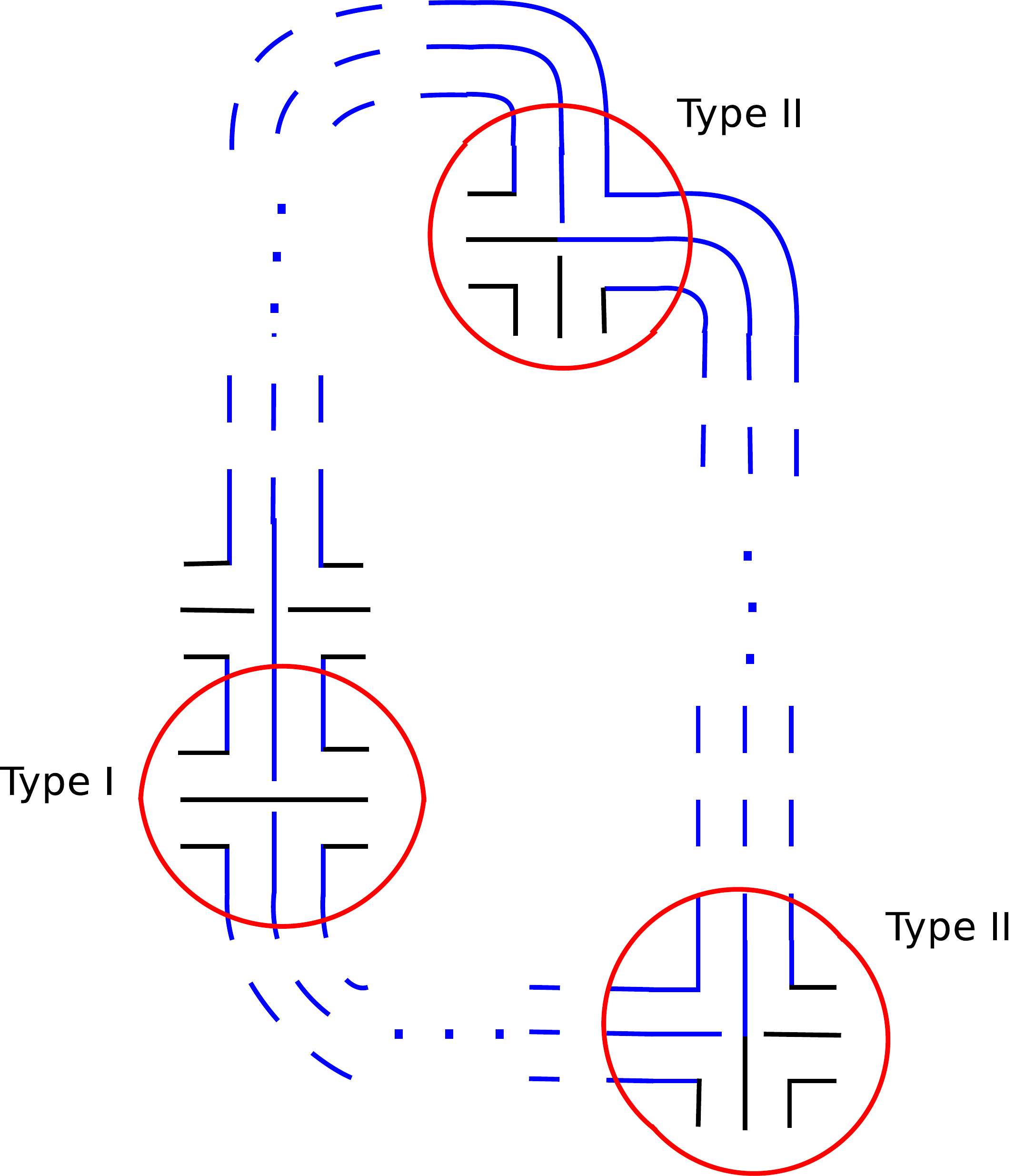}
\caption{\label{oddcycle} Form of one odd cycle of a non bipartite graph once one has cut the edges which do not belong to the cycle.}
\end{minipage}
\hspace{0.5cm}
\begin{minipage}[b]{0.45\linewidth}
\centering
\includegraphics[scale=0.34]{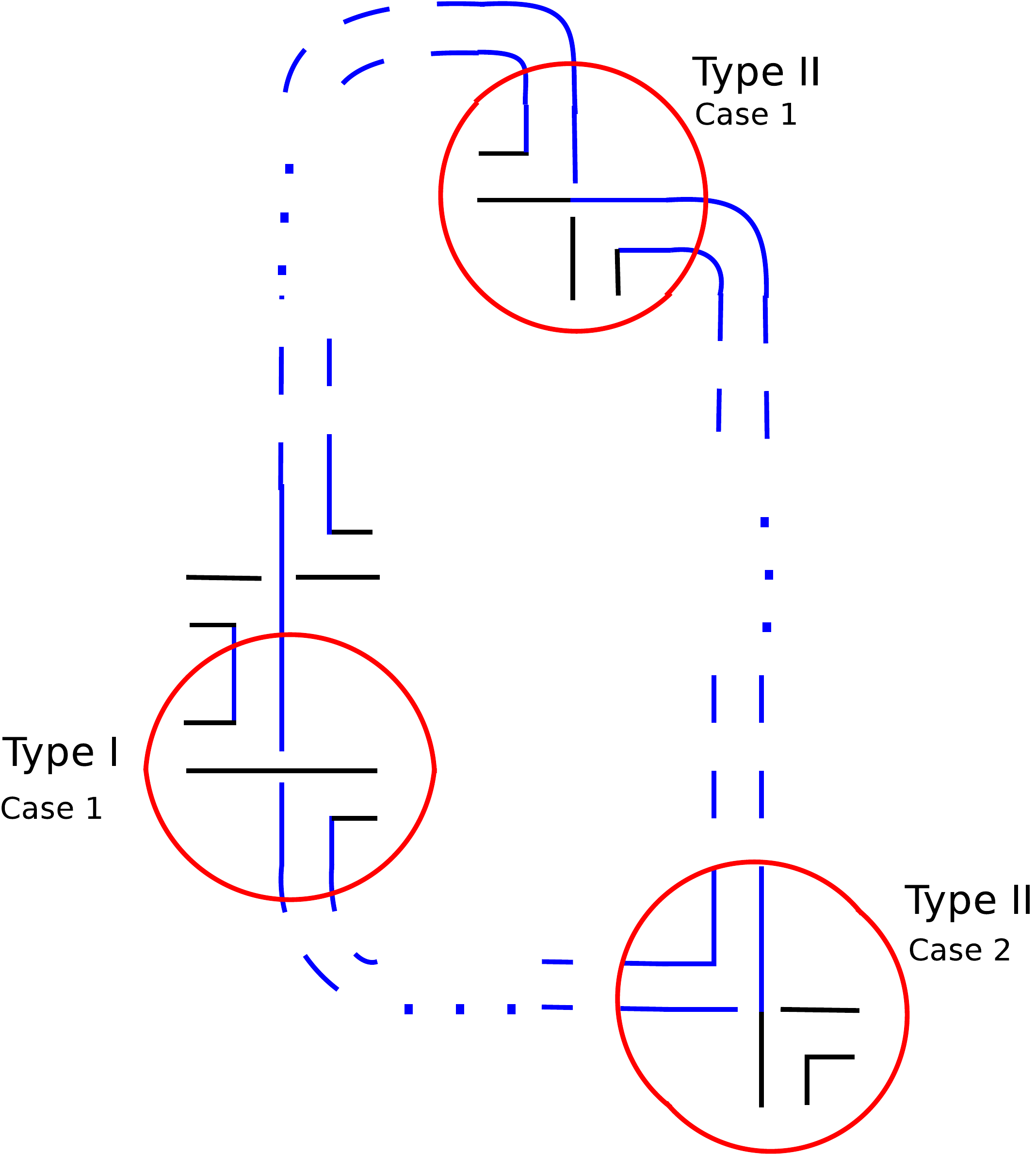}
\caption{\label{situationI-II} Different possibilities that can arise in the type $II$ vertices when one chooses a jacket with the central strand.  }
\end{minipage}
\end{figure}

Firstly we cut all the edges adjacent to any vertex of the cycle, but which do not belong to the cycle, and replace them by pairs of half-edges.
We distinguish two types of vertices:
\begin{enumerate}
\item The first type $I$ corresponds to vertices with opposite half-edges 
\item The second type $II$ corresponds to vertices with adjacent half-edges (i.e. carrying different labels $(+)$ and $(-)$ at the vertex).
\end{enumerate}
We first notice that the number of type $I$ vertices must be even. This is a direct consequence of the multi-orientability of the graph. 
We obtain a cycle of the form of the Fig. \ref{oddcycle}.  We know that the outer jacket is orientable hence we search for a non orientable jacket of type either $\bar a$ or $\bar b$. In such jackets
every vertex is twisted. 
Suppose the jacket is of type $\bar a$. It contains two corner strands of type $b$. Either each such corner strand contains one half-edge (case $1$) or one contains two half-edges and the other none (case $2$). Fig. \ref{situationI-II} shows all these possibilities. Notice that a type $I$ vertex is always in the case $1$. Moreover it can be checked that if a type $II$ vertex is case $1$ for $\bar a$ then it is case $2$ for $\bar b$.

Each type $I$ vertex, when untwisted, introduces one twist on one of its adjacent edges. A vertex of type $II$, when untwisted, introduces on its adjacent edges one twist in case $1$ and none in case $2$. An odd total number of twists along the cycle implies that the jacket is non orientable. Since the number of type $I$ vertices is even, the number of type $II$ vertices is odd. Hence either jacket $\bar a$ or $\bar b$ has an odd number of type $I$ case $1$ vertices (as case $1$ and $2$ are exchanged when $\bar a$ and $\bar b$ are exchanged) and that jacket is non orientable.  As long as one has found one non orientable jacket (as it can be done in following the steps described above) one knows that the degree of the graph is at least $\frac{1}{2}$. \qed



\bigskip

Let us now state the following:

\begin{proposition}
 If $\mathcal{G}$ is a bipartite, vacuum graph of degree zero, then $\mathcal{G}$ has a face with two vertices. 
\end{proposition}

\textit{Proof.}
Let's call $\mathcal{F}_p$ the number of faces of length $p$, $l_{\rho}$ the length of the $\rho^{\mbox{th}}$ face. 
We then have the following identities:
\begin{align}
 \label{eqfaces}
 \sum_{p\ge 1} \mathcal{F}_p&=\mathcal{F}_{\mathcal{G}}=\frac{3}{2}\mathcal{V}_{\mathcal{G}}+3, \\
 \sum_{\rho}l_{\rho}&=\sum_{p\ge1}p\mathcal{F}_p=6\mathcal{V}_{\mathcal{G}}.
\end{align}
Let us recall here that for bipartite graphs there cannot be faces of length one. 
We now write equation \eqref{eqfaces} as: 
\begin{align}
 4\mathcal{F}_2+4\sum_{p\ge3}\mathcal{F}_p&=6\mathcal{V}_{\mathcal{G}}+12, \\
 2\mathcal{F}_2+\sum_{p\ge3}p \mathcal{F}_p&=6\mathcal{V}_{\mathcal{G}}.
\end{align}
We then get by substracting the second line from the first one: 
\begin{align}
 2\mathcal{F}_2=12+\sum_{p\ge3} (p-4)\mathcal{F}_p.
\end{align}
Since 
$\mathcal{F}_3=0$, this implies that $\mathcal{F}_2 > 0$. \qed.

\medskip

We notice that this lemma remains true for non bipartite graphs but with odd cycles of minimum length 5.

\begin{proposition}
 If $\mathcal{G}$ is null degree bipartite vacuum graph, then it contains a three-edge colored subgraph with exactly two vertices.
\end{proposition}

\begin{figure}[htb]
\begin{center}
\includegraphics[scale=0.17]{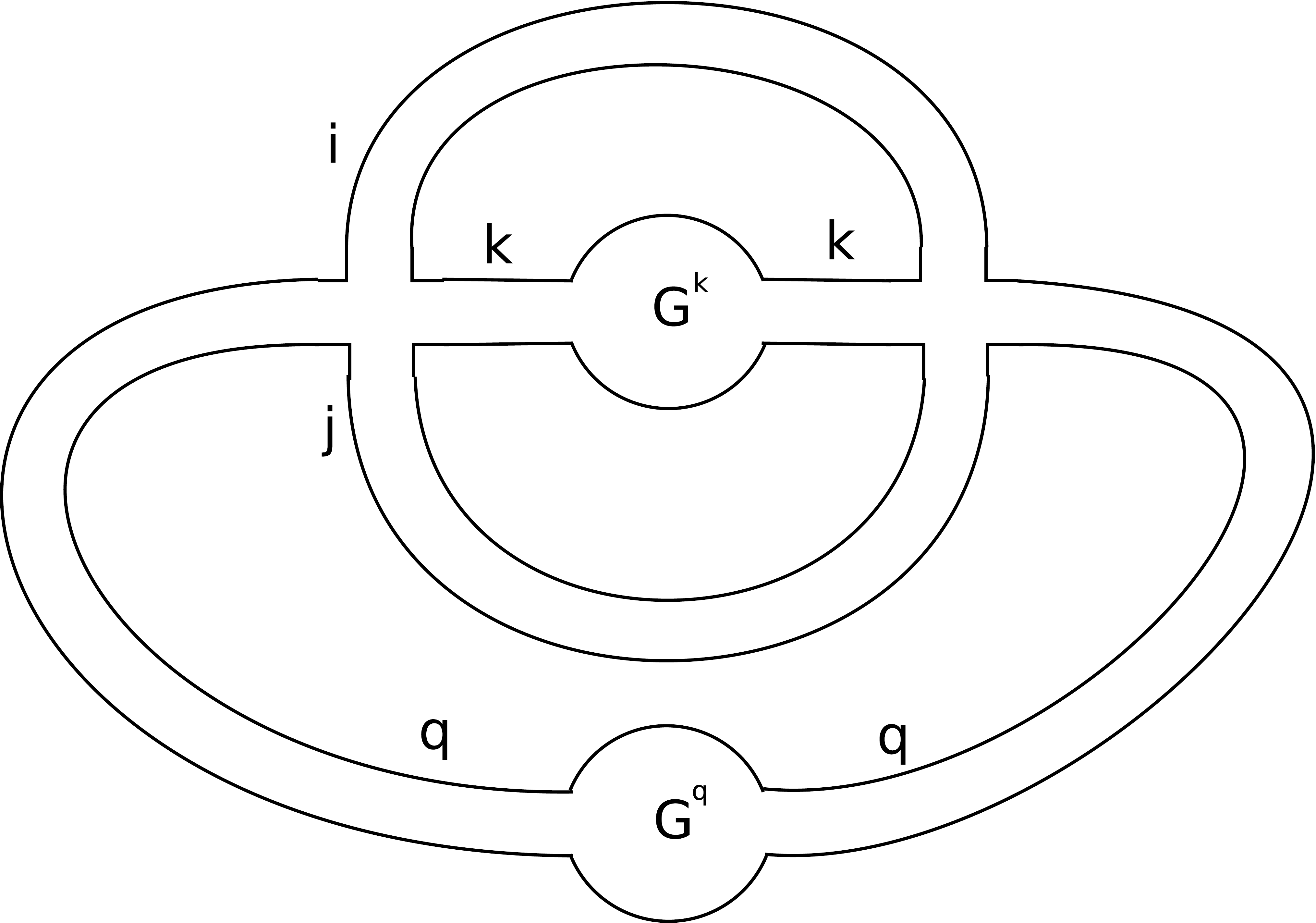}
\caption{\label{Jform} Form of the jacket not containing the face with two vertices}
\end{center}
\end{figure}

\begin{figure}[htb]
 \begin{center}
  \includegraphics[scale=0.17]{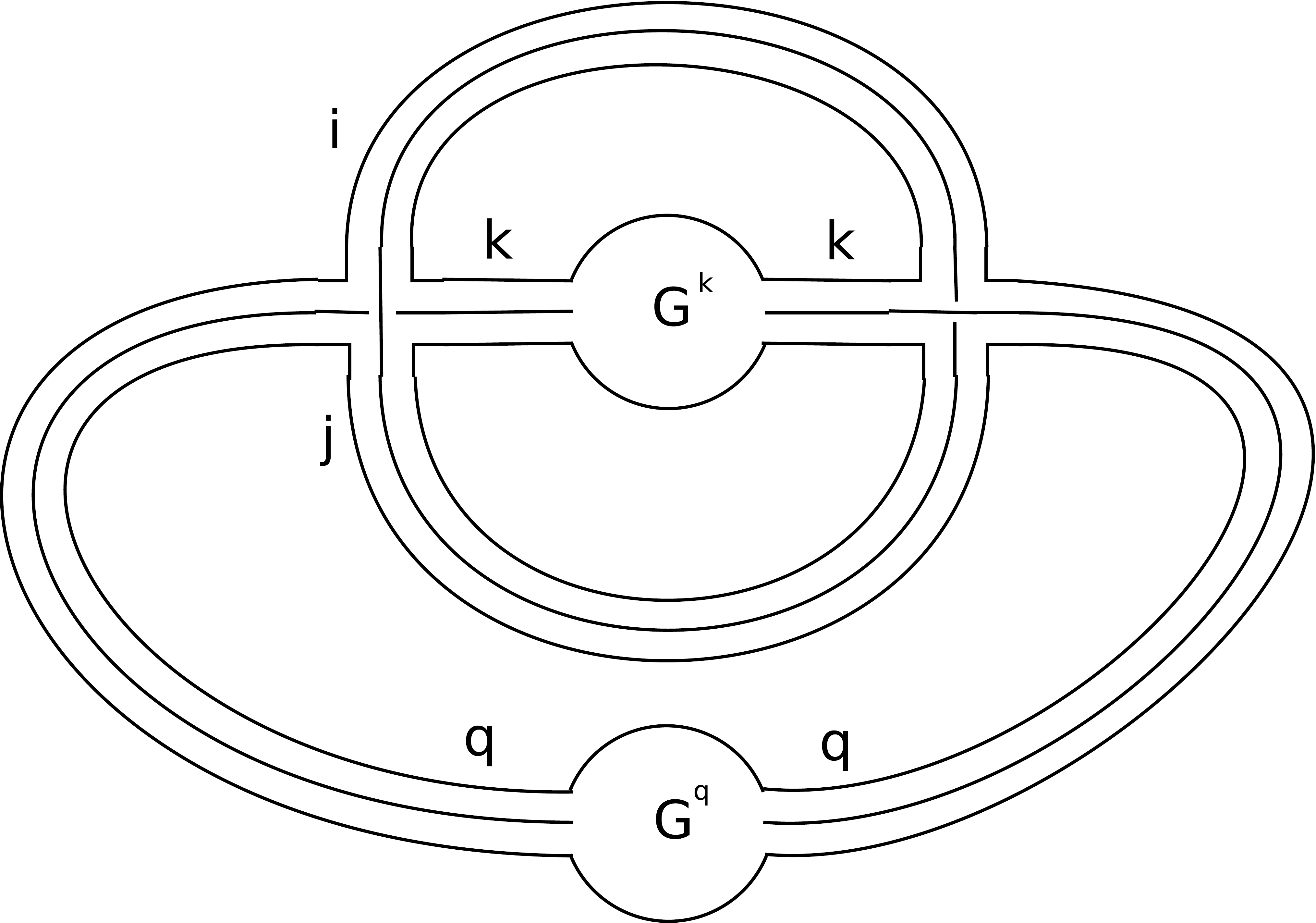}
  \caption{\label{Gform} Form of the graph $\mathcal{G}$.}
 \end{center}
\end{figure}

\textsl{Proof.}
From the previous Proposition we know that $\mathcal{G}$ has a face with two vertices, face which we denote by  $f_1$. Since $\mathcal{G}$ is bipartite, we can choose a four-coloration of its edges.
Once we have chosen a coloration, $f_1$ has two colors, denoted by  $i$ and $j$. 
We then consider the jacket $\mathcal{J}$ not containing the face $f_1$. 
Since $\varpi(\mathcal{G})=0$
the jacket $\mathcal{J}$ is planar and of the form of Fig.\ref{Jform}.
We delete the two lines of color $k$ and we get a 
new ribbon graph $\mathcal{J}'$. Since the number of vertices and faces does not change, and the number of edges decreases by two, we have
\begin{equation}
 \chi(\mathcal{J}')=\chi(\mathcal{J})+2=4
\end{equation}
and thus $\mathcal{J}'$ has two planar components. This implies that $\mathcal{G}$ is two particle reducible for any couple of colored lines touching $i$, $j$. The graph  $\mathcal{G}$ is thus of the form
of Fig. \ref{Gform}.

\begin{figure}[htb]
\begin{center}
\includegraphics[scale=0.17]{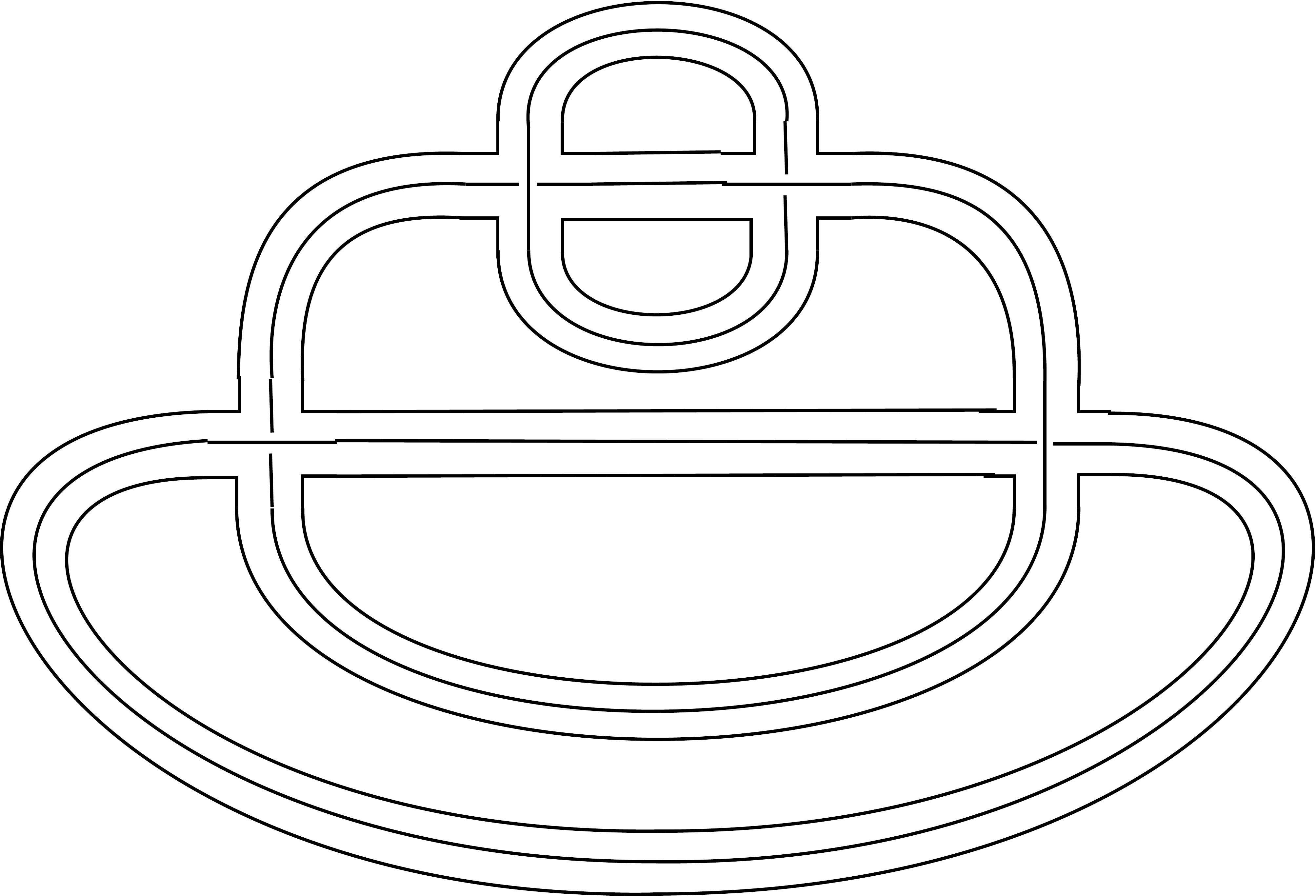}
\caption{\label{finalmelon} Form of the graph $\mathcal{G}$. It is made of successive insertion of three-edge colored subgraphs with two vertices on the lines of a "bigger" graph.}
\end{center}
\end{figure}

If the subgraph $\mathcal{G}^q$ is empty then the lines of color different of $k$ form a subgraph with three colors. Otherwise, we cut the external lines of $\mathcal{G}^q$
and reconnect the two external half lines into a new line of color $q$; we call the resulting new graph 
$\tilde{\mathcal{G}^q}$. One remarks that the bipartite caracter is conserved through this procedure.
 Moreover it is of null degree, thus from the previous Proposition it has 
a face of length two. We can thus apply the same reasoning, recursively, to this graph. 
Finally, the graph $\mathcal{G}$ is of the form of Fig. \ref{finalmelon}.
\qed

\bigskip

We have thus proved in these last two sections the main result of this paper:

\begin{theorem}
 The \iid,  \mo model \eqref{freeenergy} admits a $1/N$ expansion whose leading graphs are the melonic ones.
\end{theorem}

\section{Conclusion and perspectives}
\renewcommand{\theequation}{\thesection.\arabic{equation}}
\setcounter{equation}{0}

We have extended in this paper the $1/N$ expansion to i.i.d. \mo tensor models,
which contain a significantly larger class of graphs than 
the colored model. We have proved that the melonic graphs are still the leading ones 
in the large $N$ limit. The next steps in this direction could be:

\begin{itemize}

\item 
generalize these results to the GFT \mo model with Boulatov projection.

\item generalize to graphs with external edges (both in the i.i.d. and in the Boulatov case). 

\item search for an \mo analogous of the so-called ``uncoloring mechanism'' 
developed for colored models in \cite{uncolored}. 

\item define associated \mo tensor field theories (i.i.d. or Boulatov-like), and
study their properties, in particular their renormalizability 
and beta functions as has been done for 
colored models \cite{ren}.

\item study the (classical) Noether currents of \mo tensor models, 
as was done in \cite{joseph} again for colored models.

\item enlarge the \mo framework studied in this paper to include still larger classes of 
tensor graphs and check whether they admit a $1/N$ expansion. 

\end{itemize}

Concerning this last point, remark however that the three-dimensional \mo models seem to generate the 
largest class of tensor graphs which admits jackets (see the proof of Proposition \ref{jac}).


 

\section*{Acknowledgments}
The authors
are grateful to R\u azvan Gur\u au for fruitful  discussions.
A. Tanasa acknowledges the grants PN 09 37 01 02 and CNCSIS Tinere Echipe 77/04.08.2010.
V. Rivasseau acknowledges a Perimeter Institute grant and the ANR LQG09 grant.


\noindent
{\small ${}^{a}${\it Universit\'e Paris 13, Sorbonne Paris Cit\'e, \\
LIPN, Institut Galil\'ee, 
CNRS UMR 7030, F-93430, Villetaneuse, France, EU}}\\
{\small ${}^{b}${\it Laboratoire de Physique Th\'eorique,
Universit\'e Paris 11,}} \\
{\small {\it 91405 Orsay Cedex, France, EU}}\\
{\small ${}^{c}${\it Perimeter Institute for Theoretical Physics, 31 Caroline St. N, ON, N2L 2Y5, 
Waterloo, Canada}}\\
{\small ${}^{d}${\it 
Horia Hulubei National Institute for Physics and Nuclear Engineering,\\
P.O.B. MG-6, 077125 Magurele, Romania, EU}}\\
\\

\end{document}